\documentclass{icrc29}
\usepackage{graphicx,amssymb,amsmath,times}
\begin{document}
\title[Search for UHECR Point Sources in HiRes Stereo Data]
{Search for Point Sources of Ultra--High-Energy Cosmic Rays Above 
$10^{19}$\,eV Using a Maximum Likelihood Ratio Test}
\author[HiRes Collaboration] {S. Westerhoff$^a$, C.B. Finley$^a$ 
for the High Resolution Fly's Eye Collaboration\\
(a) Columbia University, Department of Physics, New York, NY 10027, USA }
\presenter{Presenter: C.B. Finley (finley@phys.columbia.edu), \ 
usa-westerhoff-S-abs1-he14-oral}

\maketitle

\begin{abstract}

We present the results of a search for cosmic ray point
sources at energies above $1.0\times 10^{19}$\,eV in the
HiRes stereo data set.
The analysis is based on a maximum likelihood ratio test
using the probability density function for each event rather
than requiring an a priori choice of a fixed angular bin size.
The search is extended to the combined data set of HiRes data
above $1.0\times 10^{19}$\,eV and AGASA data above
$4.0\times 10^{19}$\,eV.  In both cases,
no statistically significant clustering of events consistent
with a point source is found.

\end{abstract}

\section{Introduction}

When analyzing skymaps of cosmic ray arrival directions, it is important to correctly
account for the error on the angular resolution of individual events.  This is 
especially true in cases where data from detectors with different angular resolution
are combined, or when the angular resolution of the detector varies strongly with energy
or event geometry.  A method that correctly accounts for the angular resolution on an
event-by-event basis was recently introduced in~\cite{apj2005}.  In this paper, we
perform a likelihood ratio test of the hypothesis that several events in the 
skymap come from a common source.  In other words, we test whether any given 
position on the sky harbors a source which contributes $n_{s}\geq 1$ 
source events to the data set.  The likelihood of this hypothesis is compared
to the null hypothesis $n_{s}=0$ and this likelihood ratio is maximized using 
$n_{s}$ as a free parameter.  By calculating the likelihood ratio for a 
dense grid of points on the sky, we essentially search the entire sky for the 
most likely position of a source of $n_{s}$ events.  The statistical significance
is then estimated by applying the same method to a large set of random isotropic 
data sets and evaluating what fraction of them have a likelihood ratio which 
is equal to or larger than the ratio observed in the real data.

In~\cite{apj2005}, this method is applied to the combined HiRes and AGASA data set of events
with energies in excess of $4\times 10^{19}$\,eV.  No excess is found.
The strongest signal, from the position of the AGASA ``triplet,''~\cite{agasa2003}
is consistent with random clustering even if the HiRes threshold is lowered 
{\it a posteriori} to include a HiRes event with energy $3.76\times 10^{19}$\,eV 
in the vicinity of the ``triplet.''  
Nevertheless, a point source at that position has been claimed~\cite{glennys,troizky}, 
partly based on the fact that another HiRes event with energy between $1.0\times 10^{19}$\,eV
and $3.0\times 10^{19}$\,eV is found in the same region.  Here, we extend the energy 
range of this analysis down to $1.0\times 10^{19}$\,eV to understand whether the additional low 
energy event strengthens the case for a point source or is consistent with expectations
for an isotropic skymap with 271 events, the number of events in the HiRes stereo data set
above $1.0\times 10^{19}$\,eV.  We will analyze the HiRes stereo data set alone and in combination
with the AGASA data set, with the understanding that every ``signal'' found in the combined set
would be strongly biased as the AGASA threshold of $4.0\times 10^{19}$\,eV was chosen 
in~\cite{agasa1996} precisely because it maximized the clustering signal.
Also, magnetic smearing, which is likely to become an issue for energies as low as
$1.0\times 10^{19}$\,eV, is not taken into account.

\section{Analysis}

We give a short description of the likelihood method used in this analysis. 
Consider first a fixed source location in right ascension and declination: 
$\vec{x}_{s} = (\alpha_s, \delta_s)$.  Given a sample of $N$ cosmic ray events,
we suppose that $n_{s}$ events come from the source location $\vec{x}_{s}$,
and that the remaining $N - n_{s}$ events are random background events.  
If the $i$th event is a source event, then its true arrival direction is 
$\vec{x}_{s}$.  The probability density for finding it at some location 
$\vec{x}$ is given by the function $Q_{i}(\vec{x},\vec{x}_{s})$, derived from 
the angular errors of the event and the angular displacement between 
$\vec{x}_{s}$ and $\vec{x}$.  On the other hand, if the $i$th event is a
background event, then the probability density for finding it at a location 
$\vec{x}$ is given by the function $R_{i}(\vec{x})$, derived from the relative 
exposure of the detector to an isotropic background of cosmic rays. The subscript 
is necessary because $R$ depends on whether event $i$ is a HiRes or AGASA event.
Each of these functions is normalized to unity over positions $\vec{x}$ in the sky.

We do not hypothesize which individual events are source or background
events.  We only suppose that there are $n_s$ events in the sample that 
come from some source position $\vec{x}_{s}$.
Therefore, the partial probability distribution of arrival directions $\vec{x}$
for the $i$th event is given by:
\begin{equation}
P_i(\vec{x},\vec{x}_{s}) = \frac{n_{s}}{N}\,Q_i(\vec{x},\vec{x}_{s})
                + \frac{N-n_{s}}{N}\,R_i(\vec{x})~~.
\end{equation}
It follows that the probability of finding the $i$th event at the location 
$\vec{x}_{i}$ (where it is actually observed) is $P_i(\vec{x}_{i},\vec{x}_{s})$.  

The likelihood for the entire set of $N$ events is then given by:
\begin{equation}
{\mathcal L}(n_{s},\vec{x}_{s}) = \prod_{i=1}^{N} P_{i}(\vec{x}_{i},\vec{x}_{s})~~.
\end{equation}
The best estimate for the number of source events, under the 
assumption of a point source located at $\vec{x_s}$, is 
determined by finding the value of $n_{s}$ which maximizes $\mathcal{L}$.

Because the value of the likelihood function depends on the 
number of events, a more useful quantity than $\mathcal{L}$ 
is the likelihood ratio $\mathcal{R}$:
\begin{eqnarray}
{\mathcal R}(n_{s},\vec{x}_{s}) & = & \frac{{\mathcal L}(n_{s},\vec{x}_{s})}
        {{\mathcal L}(0,\vec{x}_{s})}\nonumber \\
    &  = & \prod_{i=1}^{N}~
        \left\{\frac{n_{s}}{N}
           \left(\frac{Q_{i}(\vec{x}_{i},\vec{x}_{s})}
                      {R_{i}(\vec{x}_{i})}
           -1\right)
        +1\right\}
\end{eqnarray}
where ${\mathcal L}(0,\vec{x}_{s})$ is the likelihood function of the 
{\it null hypothesis} ($n_{s}=0$).  In practice, we maximize $\ln\mathcal R$,
which is equivalent to maximizing $\mathcal{L}$.

The method described so far is sufficient for testing a specific source
position $\vec{x}_{s}$.  To search the entire sky for the source position
with the strongest signal, we
calculate the likelihood ratio $\ln\mathcal R$ for a dense grid of points 
on the sky covering the full range of equatorial coordinates accessible 
to AGASA and HiRes.  The source position is
essentially treated as a free parameter, along with the number
of source events $n_{s}$.  Searching for the parameters
$\alpha_{s}$, $\delta_{s}$, and $n_{s}$ which maximize the likelihood
ratio will therefore give us the best estimate for the 
position of the source and the number of events it contributes.

The search proceeds as follows.  The visible region of the sky
is divided into a fine grid of points with separations 
$0.1^{\circ}/\cos\delta$ and $0.1^{\circ}$ in $\alpha$ and $\delta$, 
respectively.  Each point is treated in turn as a source location
$\vec{x}_{s}$.  At each point, the specific quantities
$Q_{i}(\vec{x}_{i},\vec{x}_{s})$ and $R_{i}(\vec{x}_{i})$ are
required for every event.
The source probability density function $Q_{i}(\vec{x}_{i},\vec{x}_{s})$ 
depends on the angular resolution associated with the $i$th event,
which in principle may depend on several quantities including
energy, zenith angle, etc.
The background probability density function $R_{i}(\vec{x}_{i})$ 
depends on the detector exposure
to different parts of the sky: it is generally the same function for all
events observed by a given detector, but may in principle be a function
of e.g.\@ energy as well.  For each event, 
$Q_{i}(\vec{x}_{i},\vec{x}_{s})$ needs
to be reevaluated at every grid point, while $R_{i}(\vec{x}_{i})$ 
needs only to be evaluated one time.

Once the values of $Q_{i}$ and $R_{i}$ are specified,
the log likelihood ratio $\ln \mathcal R$ is maximized with respect to $n_s$,
where $n_{s}\geq 0$.
This process is repeated for each position $\vec{x}_{s}$ on the grid.
For most locations, $\ln\mathcal R$ is zero, but 
local maxima will occur in the vicinity of one or more events.  
Note that while 
the method is in some sense binned due to the discrete array of grid points, 
the spacing is chosen small enough so that errors 
introduced by binning are insignificant.

\begin{figure}[ht]
\begin{center}
\includegraphics*[width=0.9\textwidth,angle=0,clip]{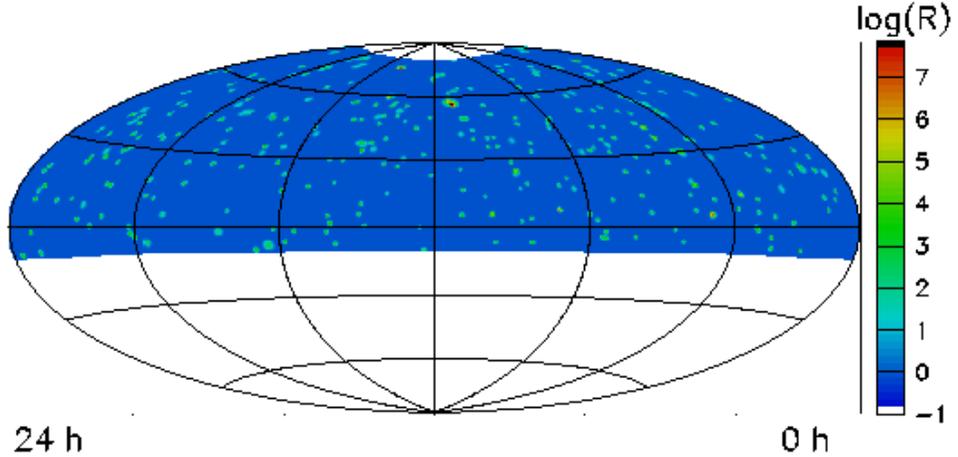}
\caption{\label {skymap_10} Likelihood ratio $\ln\mathcal R$, maximized with
respect to $n_{s}$, as a function of right ascension and declination for the 
combined set of HiRes events above $1.0\times 10^{19}$\,eV and AGASA events
above $4.0\times 10^{19}$\,eV.}
\end{center}
\end{figure}

For the signal probability density function (Q) of the HiRes events, we use a 
two-dimensional Gaussian function whose width is chosen such that $68\,\%$ of the 
probability density function falls within an opening angle $0.6^{\circ}$.
Note that for a two-dimensional Gaussian distribution the opening angle 
$\theta=1.515\,\sigma$ encloses $68\,\%$ of the distribution.  Since the 
dependence on energy is weak, we use the same value, $\sigma=0.4^{\circ}$, for
every HiRes stereo event.  The procedure to obtain the background expectation
is described in~\cite{apj2005}.

\section{Results}

We apply this method to the 271 HiRes stereo events with energies above
$1.0\times 10^{19}$\,eV.
The point with the largest $\ln\mathcal R$ is at right ascension 
$\alpha=219.2^{\circ}$ and declination $\delta=74.9^{\circ}$.  
The maximum likelihood ratio at this position is $\ln\mathcal R=8.45$ 
for $n_{s}=2.9$.  

The statistical significance of the appearance of a ``source'' with a 
maximum likelihood ratio $\ln\mathcal R$ can be evaluated using simulated 
random data sets.  The full likelihood analysis is performed for a large 
number of random data sets with the same number of events and the same 
underlying exposure as the original data set, but isotropic arrival 
directions.  The chance probability for the ``source'' to appear is then 
given by the fraction of random data sets which have at least one location 
causing the maximum $\ln\mathcal R$ to be equal or larger than $8.45$, 
the value of the maximum in the real data.  For the HiRes stereo data above
$1.0\times 10^{19}$\,eV, the chance probability of the source hypothesis is
of the order of $28\,\%$.  Consequently, there is no
statistically significant evidence for clustering consistent
with a point source in the data set.

We now include
the published AGASA data set above $4.0\times 10^{19}$\,eV in this analysis.
For AGASA, we approximate the probability density by the sum of two
Gaussian functions chosen such
that the $68\,\%$ and $90\,\%$ opening angle given in~\cite{agasa1999}
is correctly reproduced.  Details can be found in~\cite{apj2005,stokes2004}.

Fig.\,\ref{skymap_10} shows the result of the analysis.  At each
$\alpha$ and $\delta$, the likelihood ratio is shown for the
number of source events $n_{s}$ which maximizes $\ln\mathcal R$.
The point with the largest $\ln\mathcal R$ is at right ascension
$\alpha=169.0^{\circ}$ and declination $\delta=56.2^{\circ}$, which is indeed the position
of the point source claimed in~\cite{glennys}.
The maximum likelihood ratio at this position is $\ln\mathcal R=8.38$
for $n_{s}=3.6$.
As before, we evaluate the chance probability for the appearance
of a source with maximum $\ln\mathcal R=8.38$ or higher in this data set 
by analyzing a large number of simulated isotropic data set, now 
containing 57 AGASA events in addition to the HiRes stereo events.  
The chance probability is $43\,\%$, so again, no evidence for clustering
consistent with a point source is found.

The HiRes experiment is scheduled to 
take data until March 2006, and the search for cosmic ray point sources at 
the highest energies will continue when more data becomes available.

\section{Acknowledgments}

This work is supported by US NSF grants PHY-9321949,                                                                
PHY-9322298, PHY-9904048, PHY-9974537, PHY-0098826,                                                                 
PHY-0140688, PHY-0245428, PHY-0305516, PHY-0307098,                                                                 
and by the DOE grant FG03-92ER40732. We gratefully                                                                  
acknowledge the contributions from the technical                                                                    
staffs of our home institutions. The cooperation of                                                                 
Colonels E.~Fischer and G.~Harter, the US Army, and                                                                 
the Dugway Proving Ground staff is greatly appreciated.

\end{document}